\documentclass[aps,prb,twocolumn,groupaddress,superscriptaddress,showpacs,floatfix]{revtex4}
\usepackage{graphicx}
\begin{document}
\title{Aharonov-Bohm Oscillation and Chirality Effect in
 Optical Activity of Single Wall Carbon Nanotubes}

\author{Fei Ye}
\affiliation{Center for Advanced Study, Tsinghua University, Beijing 100084,
China}
\author{Bing-Shen Wang}
\affiliation{National Laboratory of Semiconductor Superlattice and
Microstructure
\\ and
Institute of Semiconductor, Academia Sinica, Beijing 100083, China\\}
\author{Zhao-Bin Su}
\affiliation{Institute of Theoretical Physics, Academia Sinica, Beijing 100080,
China } \affiliation{Center for Advanced Study, Tsinghua University, Beijing
100084, China}

\begin{abstract}
We study the Aharonov-Bohm effect in the optical phenomena of
single wall carbon nanotubes (SWCN) and also their chirality
dependence. Specially, we consider the natural optical activity as
a proper observable and derive it's general expression based on a
comprehensive symmetry analysis, which reveals the interplay
between the enclosed magnetic flux and the tubule chirality for
arbitrary chiral SWCN. A quantitative result for this optical
property is given by a gauge invariant tight-binding approximation
calculation to stimulate experimental measurements.
\end{abstract}
\pacs{61.46.+w,73.22.-f,78.67.Ch} \maketitle

Aharonov-Bohm (AB) effect manifests the significance of the global
nature of the vector potential in quantum theory \cite{AB,Berry}.
Such a geometric phase effect will result in a conductance
oscillation with respect to the enclosed flux in the cylindrical
conductor. This phenomenon is known as the Altshuler-Aronov-Spivak
effect \cite{AAS}. Since the discovery of carbon nanotube
\cite{Ijima}, it provides an ideal hollow cylindrical lattice
sheet with distinguished chiral structures which stimulates
extensive studies \cite{tasaki,reich,saito2,ivchenko,mele} in
recent years. Moreover it provides a novel system for studying the
interplay between the enclosed AB magnetic flux versus the chiral
symmetry of the tubule. When an applied magnetic flux is threaded
through, it has been shown elegantly in the tight binding
approximation (TBA) (or effective mass approximation)
\cite{ajiki,Roche1,lu} that the fundamental gap of a single wall
carbon nanotube (SWCN) is a periodic function of magnetic flux.
This result is also referred to a kind of AB effect. The
corresponding spectrum feature has been further analyzed in Ref.
\cite{Roche1} where the periodicity of Van Hove singularities with
respect to the magnetic flux was also addressed \cite{Roche1}.
Experimental transport studies \cite{Bachtold, fujiwara,science}
at the finite temperature, for multi-layer carbon nanotubes, also
show consistently a current oscillation with a single flux quantum
$\phi_0$.

Since the SWCN may behave as either an insulator or a metallic
conductor depending on its chirality as well as the strength of
penetrating magnetic flux, the conventional transport study  might
not be appropriate for exploring the generic chirality dependence
of the AB effect specially for the insulating cases. Actually, as
long as the AB flux appears in the hollow SWCN, all electronic
canonical momenta will acquire a corresponding vector potential.
As a result, the exponential phase factor associated with each C-C
link induced by the threading magnetic flux together with the
orientation of the chiral trident will make for an interplay
between the threading flux and chirality. Therefore the AB-type
effect should appear in a variety of phenomena such as the optical
properties other than the transport measurements.

In this paper, we show the AB oscillation in the natural optical
activity (or natural gyrotropy) of the SWCN with arbitrary
chirality. We derive a generic expression for the natural
gyrotropy based upon a systematic symmetry analysis and further
complement a gauge invariant TBA calculation. It exhibits the
interplay among the light polarization, tubule chirality and
threaded magnetic flux and reveals the characteristic role of the
chiral index \cite{mele2}.

\begin{figure}[tbh]
\centering
\begin{minipage}[c]{0.5\textwidth}
\scalebox{0.7}[0.7]{\includegraphics{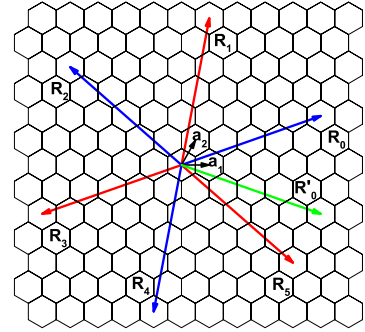}}
\end{minipage}
\caption{For given ${\bf R}_{0}[n_{1},n_{2}]$, there are five chiral vectors,
${\bf R}_{1}[n_{1}+n_{2},-n_{1}]$, ${\bf R}_{2}[n_{2},-n_{1}-n_{2}]$,
 ${\bf R}_{3}[-n_{1},-n_{2}]$, ${\bf R}_{4}[-n_{1}-n_{2},n_{1}]$ and
 ${\bf R}_{5}[-n_{2},n_{1}+n_{2}]$, denoting exactly the same SWCN
with their chiral angle deferred ${\pi}/{3}$ from each other successively. The
tubule described by ${\bf R}'_0[n_1+n_2,-n_2]$ is the mirror image of that
described by ${\bf R}_{0}$. ${\bf a}_1$ and ${\bf a}_2$ are 2D graphite lattice
basis vectors.}
\end{figure}

An ideal SWCN can be viewed as a graphite sheet rolled up into a
tubule in a fixed way along the chiral vector ${\bf R}=n_1{\bf
a}_1+n_2{\bf a}_2$ with ${\bf a}_1$ and ${\bf a}_2$ being the 2D
graphite lattice basis vectors. Its geometric structure is hence
described by a pair of integers $[n_1,n_2]$. For given
$[n_1,n_2]$, there are six chiral vectors, i.e.,~${\bf
R}_0[n_{1},n_{2}]$, ${\bf R}_1[n_{1}+n_{2},-n_{1}]$, $\ldots$,
${\bf R}_5[-n_{2},n_{1}+n_{2}]$ as shown in Fig.~1, denoting
exactly the same SWCN with chiral angles (the angle between ${\bf
R}$ and ${\bf a}_1$) as $\theta$, $\theta+\pi/3$, $\ldots$,
$\theta+{5\pi/3}$ respectively. If we introduce the chiral index
$\nu\equiv\text{mod}[n_1-n_2,3]$ as in Ref.\cite{mele2}, we find
that they can be divided into two subsets, one is $\{{\bf R}_0,
{\bf R}_2, {\bf R}_4\}$ with the same $\nu$ and the other is
$\{{\bf R}_1, {\bf R}_3, {\bf R}_5\}$ with an opposite $\nu$. The
essence of this obvious but nontrivial property, the same SWCN
with opposite chiral indices, lies in the sameness of the two
in-cell carbon atoms A and B. Actually this is the heritage to the
carbon tubule inherited from the $6mm$ symmetry of the graphite
sheet which has the $3m$ symmetry as its invariant subgroup with
corresponding quotient groups as $E$ and $C_2$(or $\sigma$). $C_2$
(or $\sigma$) will reverse the A and B atoms and also the sign of
chiral index. As a direct result of the above analysis, the
physical quantities of SWCN can be written as a periodical
function of $\theta$ with period $2\pi/3$ for $\nu=\pm 1$ and
$\pi/3$ for $\nu=0$, which are the consequences of the $3m$
symmetry and $6mm$ symmetry, respectively. When performing Fourier
expansion for these physical quantities with respect to $\theta$,
all the expansion coefficients should be functions of those
invariant quantities such as the length of chiral vector $\Lambda
|{\bf a}_1|$ ($\Lambda=\sqrt{n_1^{2}+n_2^{2}+ n_1 n_2}$) and the
greatest common divisor of $n_1$ and $n_2$ (denoted by $N$).

Since the rotation of the linear polarized light travelling
through SWCN along the tubule axis carries the information of its
chiral structure and the enclosed magnetic flux, we consider the
optical rotation power(ORP), i.e., the rotation angle per unit
length, as an observable to investigate the interplay between
chirality and threaded magnetic flux for the SWCN. By fixing the
direction of the incident light with frequency $\omega$ parallel
to the tubule axis ($z$-axis), the ORP has the expression as
\cite{landau}
\begin{eqnarray}
\chi\equiv\omega^2g_{xyz}/(2c^2)
\end{eqnarray}
where $c$ is the light speed and the third rank tensor $g_{xyz}$
is the derivative of the $xy$ component of dielectric tensor with
respected to $q_{z}$ in the long wavelength limit ${\bf
q}\rightarrow 0$, i.e., $g_{xyz}=\partial\varepsilon_{xy}
/(i\partial q_z)|_{q\rightarrow 0}$\cite{landau}. Notice that a
SWCN and its mirror image with respect to the horizontal plane
possess the same chiral index, but different chiral angle $\theta$
and $-\theta$, respectively (see Fig.~1). Further taking
consideration of the third rank tensor properties, one can see
that $\chi^{\nu}$ should be an odd function of $\theta$. Hence, it
has a Fourier series with only odd parity sine functions
\begin{eqnarray}
\chi^{\pm}(\theta)&=&a^{\pm}_1\sin(3\theta)+a^{\pm}_2\sin(6\theta)+
a^{\pm}_3 \sin(9\theta)+\cdots \; ,\nonumber\\
\chi^{0}(\theta)&=&a^0_2\sin(6\theta)+a^0_4\sin(12\theta)+\cdots
\; ,\label{fourier}
\end{eqnarray}
where the superscripts $\pm$ or $0$ denote the chiral index, and
the coefficients $a^{\nu}_n$ depend on the curvature, the magnetic
flux and the frequency of incident light. As a natural consequence
of Eq.~(\ref{fourier}) the ORP of zigzag ($\theta=0$) and armchair
($\theta=\pi/6$) tubules is zero. Moreover, considering the
rotation of $\bf R$ by $\pi/3$, since ${\bf R}_0$ and ${\bf R}_1$
correspond to the same SWCN with opposite $\nu$, it is
straightforward to see
$\chi^{\pm}(\theta+\pi/3)=\chi^{\mp}(\theta)$. Hence, we have
\begin{eqnarray}
a^{\pm}_n(\omega,\Lambda;\phi)= (-1)^{n}
a^{\mp}_n(\omega,\Lambda;\phi)\;. \label{a}
\end{eqnarray}

 We further adopt the scheme developed by White {\it et al.}
 \cite{white}, which exhibits the chiral structure of SWCN
explicitly with cylindrical geometry being properly built in. In
this scheme, all the lattice sites can be generated by repeating
the pure rotation $C_N$ and the screw operation $S(\alpha ,h)$,
while the latter one is shifting along the tubule axis by $h$ with
a simultaneous rotation around the tubule axis by $\alpha$. $h$
and $\alpha$ can be obtained through equation $h=\sqrt{3}N|{\bf
a}_1|/{2\Lambda}$, $\alpha=(2p_1n_1+2p_2n_2+p_1n_2+p_2n_1)\pi
/{\Lambda ^2}$ with integers $p_1,p_2$ satisfying
$p_2n_1-p_1n_2=N$. Hence, the Bloch momenta $\kappa\in[-\pi,\pi)$
and $n=0,1,\cdots,N-1$ as good quantum numbers can be extracted
from the characters of the $U(1)$-representations of $S(\alpha,h)$
and $C_N$, respectively. If the magnetic flux is threaded through
the SWCN, the space displacement groups $S(\alpha,h)$ and $C_{N}$
will be replaced by the corresponding magnetic displacement
groups, which are again commutable to the Hamiltonian of the flux
threaded SWCN. The Bloch momenta then become $n+\phi/\phi_0$ and
$\kappa +\alpha({\phi/\phi_0})$, respectively. This provides the
generic way of the flux dependence for all the gauge invariant
energy spectra \cite{TBA} as well as matrix elements of physical
quantity. As a result, the flux dependence of the observable for
intrinsic SWCNs will exhibit an AB oscillation with a flux period
$\phi_0$. This can be easily seen from the fact that, these
quantities can be often expressed as a double summations over
$\kappa$ and $n$, the former sums over $[-\pi,\pi)$ and then
smears the flux dependence via $\kappa+\alpha({\phi/\phi_0})$
while the latter summation will contribute the AB type flux
dependence. Then we conclude that $a^\nu_n$ in Eq.~(2) are
periodic functions of $\phi$
\begin{eqnarray}
a^{\nu}_n(\omega,\Lambda;\phi)=
a^{\nu}_n(\omega,\Lambda;\phi+\phi_0)\;. \label{ABoscillation}
\end{eqnarray}
Equations~(\ref{fourier}), (\ref{a}) and (\ref{ABoscillation})
complete the generic symmetry analysis for the OPR.

Now we apply the TBA to the SWCN with arbitrary given pair of
$[n_1, n_2]$, and ignore the curvature effect for convenience. The
TBA Hamiltonian for the SWCN with threading flux along the tubule
axis can be written as a $2\times 2$ matrix form in the momentum
space
\begin{eqnarray}
{\cal H}(\kappa,n)=V_0\left[
\begin{array}{cc}
 0 & \gamma^*_n(\kappa) \\
  \gamma_n(\kappa) & 0 \\
\end{array}\right]\;\label{hamiltonian}
\end{eqnarray}
with $V_0$ being the transfer integral equal to $2.6$~eV. The
off-diagonal matrix element has the form
$\gamma_n(\kappa)=1+e^{-i\beta_1}+e^{i\beta_2}$ with
$\beta_1={n_1\over N}(\kappa+\alpha{\phi\over\phi_0})-{2\pi
p_1\over N}(n+{\phi\over\phi_0})$ and $\beta_2={n_2\over
N}(\kappa+\alpha{\phi\over\phi_0})-{2\pi p_2\over
N}(n+{\phi\over\phi_0})$. Then it is straightforward to obtain the
energy spectrum $E^{(c,v)}_n (\kappa)=\pm V_0|\gamma_n(\kappa)|$
\cite{lu} and the one-particle wave function for the conducting or
valence band reads
\begin{equation}
|\kappa,n,c(v)\rangle={1\over\sqrt{2}}\left(|\kappa,n,A\rangle\pm
{\gamma_n(\kappa)\over
|\gamma_n(\kappa)|}|\kappa,n,B\rangle\right)
\;,\label{wavefunction}
\end{equation}
where the plus and minus signs correspond to the conducting band
and the valence band, respectively. The Bloch sum in
Eq.~(\ref{wavefunction}) reads
\begin{eqnarray}
|\kappa,n,s\rangle={1\over\sqrt{2MN}}\sum_{l=1}^N \sum_{m=-M}^M
e^{i\kappa m+i2nl\pi/N}|m,l,s\rangle\;,\label{blochstate}
\end{eqnarray}
with $s$ being the index of two in-cell atoms and $2Mh$ the tubule
length. In Eq.~(\ref{blochstate}) local state $|m,l,s\rangle$ of
unit cell $(m,l)$ can be obtained from the $(0,0)$ unit cell
$|0,0,s\rangle$ by $m$ successive screw operations $S(\alpha,h)$
combined with $l$ successive rotations $C_N$ as
$|m,l\rangle=T_{m,l}|0,0\rangle$ with $T_{m,l}\equiv S^{m}
(\alpha,h)C^{l}_N$.

For purpose of ORP calculation, we start from the off-diagonal
element of the dielectric tensor $\varepsilon_{xy}(q,\omega)$,
\begin{eqnarray}
\varepsilon_{xy}&=&{8\pi  e^2\over {m^*}^2\omega^2 V}\sum_{\kappa
n \sigma}\sum_{\kappa'n'
\sigma'}(f(E^{\sigma'}_{n'}(\kappa'))-f(E^{\sigma}_n(\kappa)))\nonumber\\
&\times&{\langle\kappa n \sigma|Q_x(-q)|\kappa'n'
\sigma'\rangle\langle\kappa'n' \sigma'|Q_y(q)|\kappa n
\sigma\rangle\over\hbar\omega+E^\sigma_n(\kappa)-E_{n'}^{\sigma'}(\kappa')+i\epsilon}\;,
\label{epsilon}
\end{eqnarray}
with $\sigma$ the band index $c$ or $v$. Here, $m^*$ and $e$ are
the electron mass and charge respectively. $V\equiv\rho\cdot
Mh|\bf R|^2/ (2\pi)$ is the system volume with $\rho$ as the
filling factor. At zero temperature $f(E)$ is simply the step
function and the vector operator ${\bf Q}(q)$ has the form $[{\bf
p}+e{\bf A}/c,e^{iqz}]_+/2$ whose $x$ and $y$ components transform
under symmetry operation $T_{m,l}$ as
\begin{eqnarray}
&&T^\dagger_{m,l}[Q_x(q)\pm iQ_y(q)]T_{m,l}\nonumber \\
&=&e^{imqh\pm i(m\alpha+l\pi /N)}[Q_x(q)\pm iQ_y(q)] \;.\nonumber
\end{eqnarray}
One then obtain the matrix elements of $Q_x\pm iQ_y$ between the
conducting and valence bands to be
\begin{eqnarray}
&&\langle\kappa',n',c|Q_x(q)\pm iQ_y(q)|\kappa,n,v\rangle=
\delta_{\kappa',\kappa+qh\pm\alpha}\delta_{n',n\pm1}\nonumber\\
&\times&\sum_{m,l}\exp \left[-i(\kappa\pm\alpha+qh)m-i{2(n\pm1)
\pi\over
N}l \right]\nonumber\\
&\times&\langle m,l,c|Q_x(q)\pm
 iQ_y(q)|0,0,v\rangle\;, \label{Qxy}
\end{eqnarray}
here the summation is over the nearest neighbors of the $(0,0)$
unit cell and it reflects the local geometric structure of SWCN.
The $\delta$-function in Eq.~(\ref{Qxy}) gives the selection rule
for the transverse optical transition. To keep the gauge
invariance, we transfer the calculation of the momentum matrix
element into that of the coordinate matrix element through the
commutation law ${\bf p}+e{\bf A}/c=m^*[{\bf r},H]/(i\hbar)$
 \cite{foreman} to avoid the explicit treatment of vector potential
${\bf A}$. The translational invariance is also rigorously kept in
our calculating procedure.  Based upon all the above
considerations, the  ORP can be calculated as
\begin{eqnarray}
&&\chi ={e^2\omega^2V_0^2\over \pi\rho
c^2}\sum_{n=1}^N\int_{-\pi}^{\pi}d\kappa
{W(\kappa,n)\Delta(\kappa,n)\partial_\kappa D(\kappa,n)
\over[(\hbar\omega+i\epsilon)^2-\Delta(\kappa,n)^2]^2}\;,\nonumber\\
&&W(\kappa,n)=1-\text{Re}\left[{\gamma_n(\kappa)\gamma_{n+1}^*(\kappa+
\alpha)\over |\gamma_{n}(\kappa)| |\gamma_{n+1}(\kappa+ \alpha)
|}e^{-i\varphi_{AB}}\right]\;,\nonumber\\
&&D(\kappa,n)=|\gamma_{n+1}(\kappa+ \alpha)|^2-
|\gamma_n(\kappa)|^2\;,\nonumber\\
&&\Delta(\kappa,n)=E^{(c)}_{n+1} (\kappa+\alpha)-
E^{(v)}_n(\kappa)\;,\label{gxyz}
\end{eqnarray}
where $\varphi_{AB}=\pi (n_1+n_2)\Lambda^{-2}$ is the difference
between the azimuth angles of the two atoms in one unit cell.

In the optical limit $q \rightarrow 0$ the minimum of
$\Delta(\kappa,n)$ gives an indirect band gap for the transverse
optical transition as $\Delta_g=\sqrt{3}\pi V_0/\Lambda$. This gap
sets a lower bound of the incident light wavelength as
$\lambda_c=2\hbar c\Lambda /(\sqrt{3}V_0)$, below which no optical
transverse absorption take place. \textit{Since this threshold is
independent of chirality, it provides a means to determine the
tubule diameter for all kinds of SWCNs by optical measurements. }

\begin{figure}[tbh]
\begin{minipage}[c]{0.5\textwidth}
\scalebox{0.58}[0.58]{\includegraphics{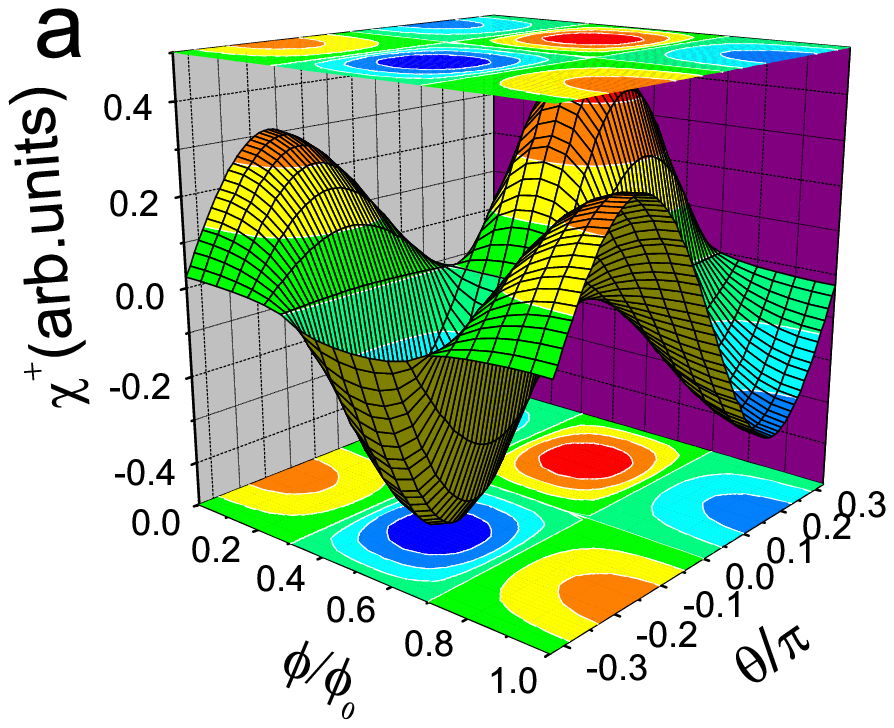}}
\scalebox{0.58}[0.58]{\includegraphics{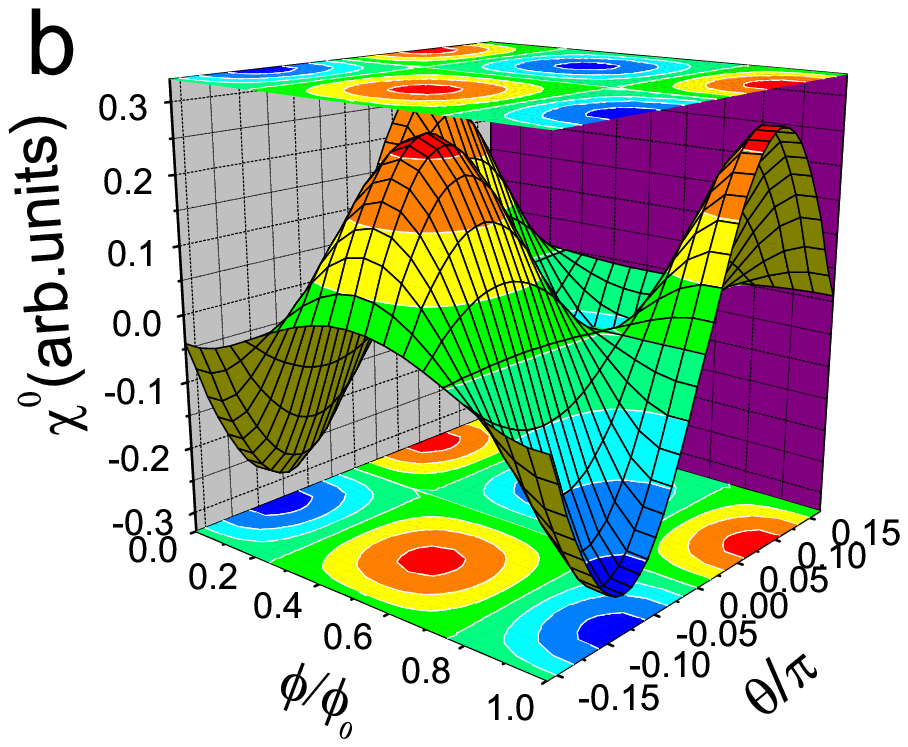}}
\end{minipage}
\caption{ The dependence of the ORP on magnetic flux and chiral
angle is plotted in (a) for $\nu=1$ and in (b) for $\nu=0$. The
normalized incident light frequency is $\bar{\omega}=0.5$ and the
tubule diameter is around $37~\text{nm}$. In (a) the ORP for
$\nu=1$ tubules is a periodic function of chiral angle with period
$2\pi/3$ and magnetic flux with period $\phi_0$. The ORP for
$\nu=-1$ tubules can be obtained through the translation of chiral
angle by $\pi/3$ in (a). In (b), the ORP for the $\nu=0$ tubules
is a periodic function of chiral angle with period $\pi/3$ and
magnetic flux with period $\phi_0$.}
\end{figure}

When plotting $\chi$ versus the chiral angle $\theta$ for
arbitrary $[n_1,n_2]$, we found the numerical data can be
interestingly regrouped into three categories in accordance with
the chiral index $\nu=0, \pm 1$. By introducing a rescaled
dimensionless frequency $\bar{\omega}=\hbar \omega /\Delta_g$,
$\chi^\nu$ can be fitted perfectly by the following functions with
parameter $f={4\omega^2e^2/\pi\rho c^2V_0}$
\begin{eqnarray}
\chi^\pm/f&\approx& {a}_1^\pm(\bar{\omega},{\phi\over\phi_0})
\sin(3\theta)+\Lambda^{-1}{a}_2^\pm(\bar{\omega},{\phi\over\phi_0})
\sin(6\theta)\nonumber\;,\\
\chi^0/f&\approx&\Lambda^{-1}{a}_2^0(\bar{\omega},{\phi\over\phi_0})
\sin(6\theta) \;,\label{orp}
\end{eqnarray}
where ${a}_1^\pm= -{a}_1^\mp$ and ${a}_2^\pm={a}_2^\mp$ which, as
dimensionless functions of $\bar{\omega}$ and $\phi/\phi_0$, can
be determined numerically.  Eq.~(\ref{orp}) is entirely consistent
with the above symmetry analysis (Eqs.~(\ref{fourier}) and
(\ref{a})). In Fig.~2 we give 3D plots of $\chi^+$ and $\chi^0$
versus magnetic flux and chiral angle. Note that $\chi^-$ can be
obtained through exact relation $\chi^-(\theta)=\chi^+
(\theta+\pi/3)$. The results explicitly shows AB oscillation in
the ORP with a single flux quantum period and $\chi^\nu$ is an
even function of $\phi$. For large $\Lambda$, the numerical data
verifies $\chi ^+=-\chi ^-$ within its precision and $\chi^0$ is
much smaller than $\chi^\pm$ in magnitude (see also
Eq.~(\ref{orp})). It is interesting to note that for $\nu=\pm 1$
tubules the magnitude of $\chi^\pm$ reaches its maximum/minimum
when chiral angle approach ${\pi}/{6}$ for fixed $\phi$ and
$\bar{\omega}$, however for the armchair tubules
($\theta={\pi}/{6}$) $\chi^0=0$. This implies that the chiral
index plays a role much more sensitive than the chiral angle in
the natural gyrotropy properties of SWCN.

\begin{figure}[tbh]
%\begin{minipage}[c]{0.5\columnwidth}
\scalebox{0.72}[0.72]{\includegraphics{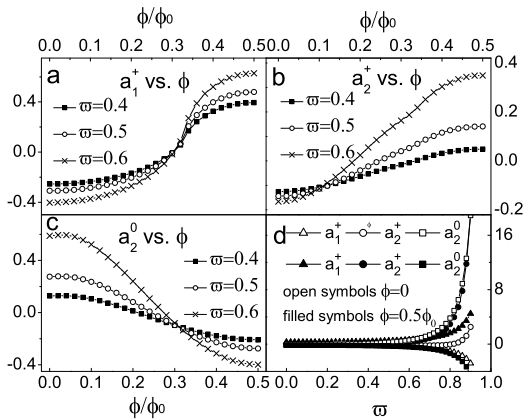}}
%\end{minipage}
\caption{ Coefficients $a_1^+$, $a_2^+$, and $a_2^0$ are plotted
as functions of magnetic flux $\phi$ for different values of
renormalized frequency $\bar{\omega}$ in (a), (b), and (c),
respectively. The filled square, open circle and cross symbol
correspond to $\bar{\omega}=0.4$, $0.5$ and $0.6$, respectively.
Fig.(d) is the plot of $a_1^+$, $a_2^+$, and $a_2^0$ vs.
$\bar{\omega}$ denoted by triangle, circle and square,
respectively, for different values of magnetic flux. The open
symbols are the data for $\phi=0$ and the filled symbols for
$\phi=0.5\phi_0$.}
\end{figure}

The fitting coefficients $a_1^+$, $a_2^+$, and $a_2^0$ are plotted
in Figs.~3a, 3b, and 3c respectively as the functions of
$\phi/\phi_0$ with $\bar{\omega}=0.4, 0.5, 0.6$. All ${a}_1^+$
with different $\bar{\omega}$ coincide approximately at a common
node ${\phi/\phi_0}=0.31$, i.e., coincide at $\pm 0.31+\mu$ with
$\mu$ being the integer, since it is a periodic even function of
$\phi/\phi_0$. However, for the $\nu =0$ tubules, the position of
the zero point changes distinctively versus $\bar{\omega}$. Since
${a}_1^+$ dominates the $\chi$ function, this remarkable fact
provides a way to distinguish experimentally whether the SWCN
under investigation belong to $\nu=\pm 1$ or $\nu=0$. In Fig.~3d
we plotted ${a}_1^+$, ${a}_2^+$, ${a}_2^0$ as functions of
$\bar{\omega}$ for fixed ${\phi/\phi_0}=0, 0.5$. It is seen that
when the incident light frequency approaches the band edge the ORP
for both two kinds ($\nu=0$ and $\nu=\pm 1$) of SWCN increases
rapidly. This fact may also be helpful to realize the ORP with
enough intensity by tuning the proper frequency of the incident
light. To have a quantitative estimation for the ORP of SWCN, we
choose tubule diameter 10~$\AA$ and the incident light wavelength
1.4~$\mu$m, $\phi=0$, and $\rho=2$, which corresponds to
$\bar{\omega}\sim0.8$ and $\Lambda\sim 13$, then the magnitude of
the rotation angle per unit length can be calculated to be
$79~\text{rad}/\text{cm} \times \nu\sin(3\theta)$ for $\nu=\pm 1$
tubules and $23~\text{rad}/\text{cm} \times\sin(6\theta)$ for
$\nu=0$ tubules.

As the final remarks, the AB effect is conventionally investigated
in connection with the transport studies for the cylindrical
metallic sheet. In this paper, we show that the AB oscillation
will also appear in the optical phenomena generically for
arbitrary chiral SWCNs. Our arguments apply even to other kinds of
chiral single wall nanotubes. Moreover, the response to the
enclosed magnetic flux is dramatically affected by the chiral
index which characterizes different kinds of global helicity
states of SWCNs and permits distinguished chirality dependence of
physical properties. The gauge invariant TBA calculation not only
provides quantitative results for various quantities to stimulate
the experimental measurements but also verifies all the
conclusions drawn from our generic symmetry analysis.


\begin{thebibliography}{}
\bibitem{AB} Y. Aharonov and D. Bohm, Phys. Rev. {\bf 115}, 485(1959).
\bibitem{Berry} M. V. Berry, Proc. R. Soc. London Ser. A {\bf
392}, 45(1984).
\bibitem{AAS} A. G. Aronov and Yu V. Shavin, Rev. Mod. Phys. {\bf 59},
755(1987).
\bibitem{Ijima}S. Iijima, Nature(London) {\bf 354}, 56(1991).
\bibitem{tasaki} S. Tasaki, K. Maekawa, and T. Yamabe, Phys. Rev.
 B {\bf 57}, 9301(1998).
\bibitem{reich} S. Reich and C. Thomsen, Phys. Rev. B {\bf 62}, 4273(2000).
\bibitem{saito2} R. Saito, G. Dresselhaus, and M. S. Dresselhaus, Phys. Rev.
 B {\bf 61}, 2981(2000).
\bibitem{ivchenko} E. L. Ivchenko and B. Spivak, Phys. Rev.
 B {\bf 66}, 155404(2002).
\bibitem{mele} N. Sai and E. J. Mele, cond-mat/0308583
\bibitem{ajiki} H. Ajiki and T. Ando, J. Phys. Soc. Jpn. {\bf 65}, 505(1996).
\bibitem{lu} J. P. Lu, Phys. Rev. Lett. {\bf 74}, 1123(1995).
\bibitem{Roche1} S. Roche, {\it et al.}, Phys. Rev. B {\bf 62}, 16092(2000).
\bibitem{Bachtold} A. Bachtold, {\it et al.}, Nature(London), {\bf 397},
673(1999).
\bibitem{fujiwara} A. Fujiwara, {\it et al.}, Phys. Rev. B {\bf 60},
13492(1999).
\bibitem{science}S. Zaric, {\it et al.}, Science {\bf 304}, 1129
(2004); U.C. Coskun, {\it et al.}, Science {\bf 304}, 1132 (2004),
these two papers are noted  thanks to the referee's kindly
reminder after our submission.
\bibitem{landau} L. D. Landau and E. M. Lifshitz, \textit{Electrodynamics
of Continuous Media}, (Pergamon Press, Oxford, 1984).
\bibitem{mele2} E. J. Mele and P. Kr$\acute{\text{a}}$l, Phys. Rev. Lett. {\bf
88}, 056803(2002).
\bibitem{white} C. T. White,  D. H. Robertson, and J. W. Mintmire, Phys. Rev. B {\bf 47}, 5485(1993).
\bibitem{TBA} The corresponding energy spectra of the flux threaded SWCN has been solved in the TBA in
Ref. [11].
\bibitem{foreman} B. A. Foreman, Phys. Rev. B {\bf 66},
165212(2002).
\end{thebibliography}
\end{document}